\documentclass[aps,prb,twocolumn,floats]{revtex4}
\usepackage{amssymb}
\usepackage{amsbsy}
\usepackage{amsmath}
\usepackage{graphicx}
\usepackage{amsmath}
\usepackage{times}
\usepackage{color}
\usepackage{setspace}
\usepackage{bm}
\newcommand\bea{\begin{eqnarray}}
\newcommand\eea{\end{eqnarray}}
\newcommand\beq{\begin{equation}}
\newcommand\eeq{\end{equation}}

\begin{document}
\title{Photo-induced Entanglement in a Magnonic Floquet Topological Insulator}

\author{Satyaki Kar}
\affiliation{{Institute of Physics, Bhubaneswar-751005, India}}
\email{satyaki.phys@gmail.com}
\author{Banasri Basu}
\affiliation{Physics and Applied Mathematics Unit, Indian Statistical Institute, Kolkata - 700108, India.}
\email{sribbasu@gmail.com}
\begin{abstract}
When irradiated via high frequency circularly polarized light, the stroboscopic dynamics in a Heisenberg spin
system on a honeycomb lattice develops a next nearest neighbor (NNN)
Dzyaloshinskii-Moriya (DM) type term\cite{owerre}, making it a magnonic Floquet topological insulator. {We investigate 
the entanglement generation and its evolution on such systems - particularly an irradiated ferromagnetic XXZ spin-$\frac{1}{2}$ model in a honeycomb lattice
as the system parameters are optically tuned. In the high frequency limit, we compute the lowest quasi-energy state entanglement in terms of the 
concurrence between nearest neighbor (NN) and NNN pair of spins and witness the entanglement transitions occurring there.
For the easy axis scenario, the unirradiated system forms a product state but entanglement grows between the NNN spin pairs 
beyond some cut-off DM strength.
Contrarily in easy planar case, NN and NNN spins remain already entangled in the unirradiated limit. It then goes through an entanglement transition which causes decrease (increase) of the NN (NNN) concurrences  down to zero (up to some higher value) at some critical finite DM interaction strength. For a high frequency of irradiation and a suitably chosen anisotropy parameter, we can vary the field strength to witness sudden death and revival of entanglement in the Floquet system.}
Both exact diagonalization and modified Lanczos techniques are used to obtain the results upto 24 site lattice.
We also calculate the thermal entanglement and obtain estimates for the threshold temperatures below which
non-zero concurrence can be expected in the system.
\end{abstract}
\maketitle

\section{Introduction}


Recently there is an upsurge of interest in realizing and utilizing quantum information 
aspects of various quantum many body systems (QMBS). Built at the interface of quantum information science, condensed matter
theory, statistical physics, quantum field theory, the study of many-body entangled states rapidly has become
a very active topic of research.
In this respect, quantum entanglement plays a crucial
role in the highly efficient quantum computation and quantum information processing\cite{vidal,horodechi}. With the rapid development of the 
experimental process on quantum control, there is a rapidly growing interest in entanglement generation.  Thus the 
quantification of entanglement has found a key place in quantum information process applications. 

On the other hand, recent study of Dirac and topological magnons
in solid-state magnetic systems\cite{onose,ren,chisnell,owerre0,fransson} 
is expected to open a challenging avenue towards magnon spintronics and magnon thermal devices. 
Since the magnons are charge-neutral quasiparticles, it is believed that the magnon quantum computing will offer a 
favorable pathway for eliminating
the difficulties posed by charged electrons\cite{adrianov,khitun} and as such 
the magnonic devices would be more efficient in quantum memory and information
storage\cite{simon,tanji,specht,wang}. 

In this context, our focus is to study the entanglement in a magnonic system that is irradiated via a strong 
periodic circularly polarised light. 
{There are many measures of entanglement, such as entanglement entropy, entanglement of formation, purity 
or negativity capturing the quantum correlation within an interacting system by different means.
Von Neumann entropy 
gives a standard measure for entanglement of pure states. But for a generic mixed state, an entanglement entropy
can give non-zero values for each of the subsystems even if there is no entanglement. So in those cases it is the entanglement
of formation that gives a true measure of entanglement.\cite{wooters} For a given purity or mixedness, with
different possible combination of pure states that the state can collapse into, it is the entanglement of 
formation that gives the minimum number of singlets required to create the mixed density of states.
This is a monotonically increasing function of an argument called concurrence C, with $0\le C\le1$, which by itself can also be regarded as a measure for entanglement in a mixed state, for example, between a pair of qubits
within a multi-qubit large system \cite{wooters}. In fact this is an entanglement monotone which is zero for separable states
and unity for Bell states (four maximally entangled 2-qubit states). For a pair of qubits, concurrences are well defined as will be described later.}
Our analysis shows their tunability in terms of the frequency 
of irradiation and provides significant control over the quantum information processing.

At the very outset, let us reiterate here that recent trend shows plenty of work on optical lattices involving
Dirac plasmons\cite{weick}, Dirac magnons\cite{balatsky} or photonic topological insulators\cite{khanikaev}
that have Dirac like bosonic spectrum. 
When dynamics is studied in such systems in presence of time-varying fields, plethora of 
exotic phenomena such as 
defect productions, dynamical freezing, dynamical phase 
transition or entanglement generation\cite{amit,ksrmp,kar1,kar2} can be expected.
Particularly for a  periodic quench,  one can use the Floquet theory\cite{eckart} for  
stroboscopic evolution\cite{nori} of the system, which results in an effective static 
Hamiltonian out of the originally dynamical system.

A Dirac system shows interesting dynamical features upon light irradiation\cite{lindner,lindner2,moessner,light,light0,light1,light2,light3,light4,debu,debu2,aniruddha}.
{An irradiated field can lead to nontrivial Floquet systems
  like Floquet topological insulators (FTI)\cite{moessner}, as can be seen,
for example in an irradiated semiconductor quantum well\cite{lindner} or a 3D topological insulator\cite{lindner2}. }
We know that a ferromagnetic Heisenberg spin-$\frac{1}{2}$ (FMHS) model with next nearest neighbor (NNN) 
Dzyaloshinskii -Moriya interaction (DMI) in a honeycomb lattice,
 under a linear spin wave approximation (LSWA), turns out to be a magnonic equivalent of the Haldane model - the famous primitive toy model to show topological
transitions\cite{pantaleon}.
{Interestingly, this can also be achieved via irradiation with high frequency circularly polarized light\cite{owerre}.
The resulting Floquet Hamiltonian develops easily tunable synthetic laser-induced NNN DMI in addition to a FMHS with modified anisotropy.}
 Within LSWA, the model behaves like a bosonic Haldane model enabling the system to emerge as
 topologically nontrivial at intermediate frequencies of the irradiation.

 In this paper, we probe the entanglement characteristics of such {Floquet model,
 born out of irradiating the spin system - both in their topologically trivial and non-trivial limits.
 In the infinite frequency limit (which is equivalent to zero DMI), the
 resulting lowest quasi-energy state is a ferromagnetic product state and hence unentangled,
unless the anisotropy is of easy planar type.
But with moderately high frequencies (but not small ones, as discussed in Appendix A for which other higher order terms from the high frequency expansion of the Floquet Hamiltonian also become relevant), the system can become entangled due to generation of the DMI term,}
as an antisymmetric DM exchange interaction can excite entanglement and teleportation fidelity\cite{zhang}
in the system.
{  This is a short-range interacting system and thus the entanglement transitions
 does not coincide with the topological transition} that occurs as soon as
DMI term is brought in. However, deep within the topological phase, system shows finite entanglement, irrespective of the value of anisotropy.

{While dealing with low temperature entanglement of these systems, we not only need information of the lowest energy state but that of low energy excitations as well. Following that, a measurement on thermal entanglement is very effective in this context. We compute thermal concurrences in our Floquet model and notify its behavior at the various low temperatures.}


For numerical computation, we use diagonalization methods like exact diagonalization (for small lattices with L=6 and 12) and a 
modified Lanczos technique\cite{dag} (for L=18 and 24) and obtain the concurrences there from.

{The article is organized as follows.
In section II, we start with the Hamiltonian formulation of the problem.
In section III, we introduce concurrences in the Floquet model and discuss briefly how to compute that.
Section IV details our results and the corresponding discussion and finally in section V, we conclude our work.}


\section{Hamiltonian Formulation}
A ferromagnetic spin-1/2 XXZ model is given as
\begin{eqnarray}
 H_J&=&-J\sum_{<i,j>}[ S_i^zS_j^z+\frac{\Delta_0}{2}(S_i^+S_j^-+h.c.)].
\end{eqnarray}
When such system is irradiated with light, the electric field (${\bf E}$) of the light interacts with the spin moments (${\bf \mu}$) yielding 
time periodic Aharonov-Casher phases\cite{casher} $\phi_{ij}=\frac{1}{\hbar c^2}\int {\bf E}\times${\boldmath $\mu~.dx_{ij}$}
between sites $i$ and $j$ in the lattice. This paves way for a Floquet analysis resulting in an
 effective static Hamiltonian for the dynamic system.
Particularly for high frequency circularly polarized irradiation with $E=E_0(cos~\omega t,sin~\omega t)$, a high frequency expansion can lead us to
a Floquet Hamiltonian given as,
\begin{eqnarray}
 H_F&=&-J\sum_{<i,j>}[ S_i^zS_j^z+\frac{{\Delta_\alpha}}{2}(S_i^+S_j^-+h.c.)]\nonumber\\
 &&+{D_\alpha(\omega)}\sum_{<<i,k>>}\nu_{ik}(S_i^xS_k^y-S_k^xS_i^y)\nonumber\\
 &=&-J\sum_{<i,j>}[ S_i^zS_j^z+\frac{{\Delta_\alpha}}{2}(S_i^+S_j^-+h.c.)]\nonumber\\
 &&+\frac{{D_\alpha(\omega)}}{2}\sum_{<<i,k>>}\nu_{ik}(iS_i^+S_k^-+h.c.)
 \label{eq2}
\end{eqnarray}
The details of the calculation can be found in the Appendix A.
Notice that the spin anisotropy gets altered from $\Delta_0$ to {$\Delta_\alpha=J_0(\alpha)\Delta_0$} thereby changing the spin anisotropy
parameter in the Floquet model. Here $J_n(\alpha)$ is $n$-th order Bessel's function of first kind with {$\alpha=\frac{g\mu_Ba E_0}
{\hbar c^2}$ (see Appendix-A for definition of the parameters).}
Furthermore, an additional NNN DMI term sets in having amplitude {$D_\alpha(\omega)=K(\alpha)/\omega$ where $K(\alpha)=\sqrt{3}\Delta_0^2J^2 J_1(\alpha)^2$.}
So for very large $\omega$, this is essentially zero and can only become significant otherwise.
This DMI term acts as a complex NNN hopping term, like in a spinless Haldane model and is the reason behind its topological 
nontriviality. Here $\nu_{ik}$ is a prefactor for hopping 
between sites $i$ and $k$ and
$\nu_{ik}=+1~(-1)$ for $i,k~\in~A(B)$ sublattice of the system.

\section{Concurrence in Floquet model}

In order to compute the concurrence of the ground state as well as low energy excitations of a system, we need the full energy spectrum of the 
problem and
we use numerical diagonalization of the Hamiltonian matrix to serve that purpose. {As we deal with a Floquet model here, we look out for the Floquet quasi-energy spectrum and particularly, the lowest quasi-energy state and concurrences there in.}
\begin{figure}
\vspace{.1in}
\centering
\includegraphics[width=.31\linewidth,height=1.3 in]{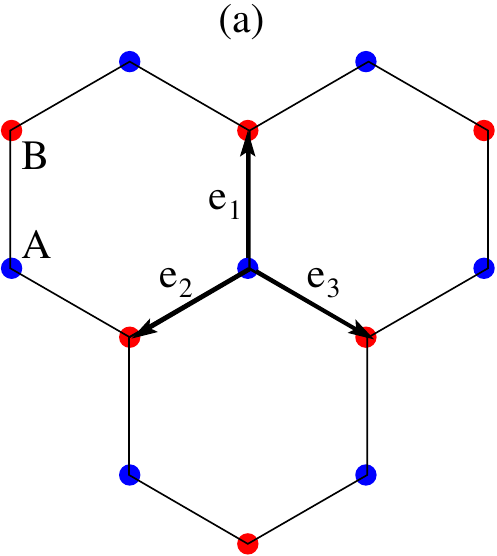}\hskip .1 in
\includegraphics[width=.64\linewidth,height=1.47 in]{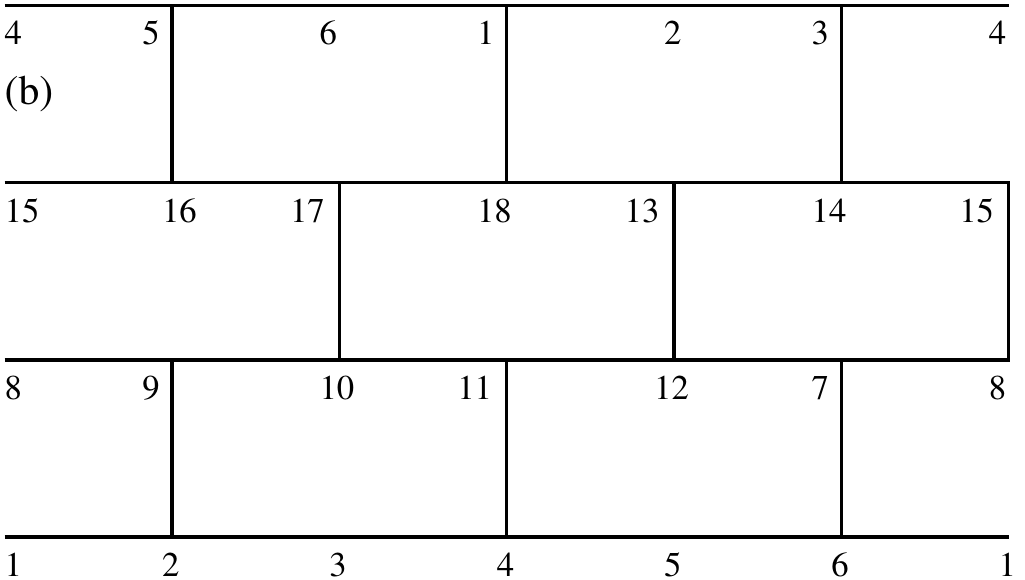}\\\vspace{.3in}
\hskip .3 in 
\includegraphics[width=.64\linewidth,height=1.27 in]{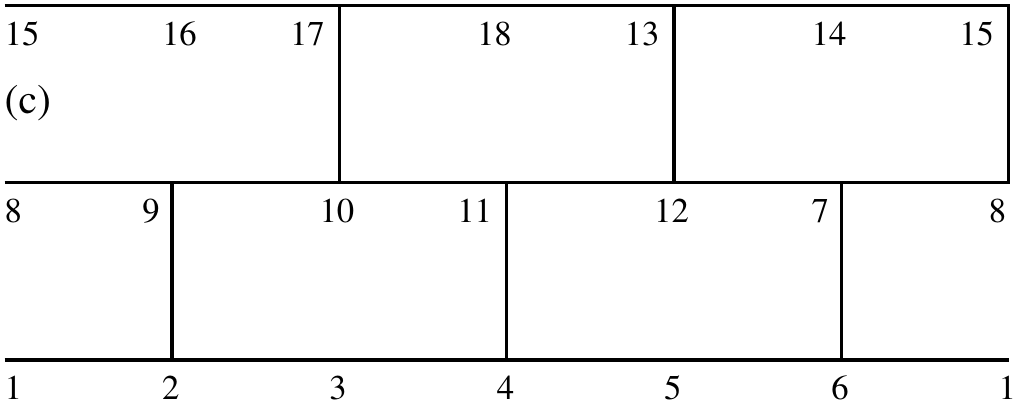}
\caption{(a) Honeycomb lattice containing sublattices A and B. Unit vectors
${\bf e_1,~e_2}$ and ${\bf e_3}$ (see Appendix A) are shown as well. Site numbering, implementing (b) PBC and (c) OBC, in a 
brickwall lattice (which is topologically equivalent to a honeycomb lattice) of size L = 18. The odd (even) numbered sites fall within the sublattice A (B).}
\label{fig1}
\end{figure}

We first briefly describe the lattice - its site numbering and its bond connections, that are necessary to
identify different interaction pairs.  A honeycomb lattice (see cartoon in Fig. 1(a)) can more 
conveniently be described using a brick-wall lattice. 
First we use periodic boundary condition (PBC) and the site numbering are given accordingly.
The example for $L=18$ site lattice can be seen in Fig.\ref{fig1}(b).
For computing concurrence between NN and NNN pairs, we considered the numbered pairs (3,4) and (3,11)
respectively.
Please note here that this numbering is not unique and we only need to ensure that the numbering and boundary conditions
do not break the symmetry of the lattice and treat each of the hexagonal plaquettes equally. {In order to see the effect on entanglement at the edges, 
later we also consider finite systems using open boundary conditions (OBC) along $y$ directions. As can be seen from Fig.\ref{fig1}(c), this amounts to pair of (zigzag) edges parallel to 
$x$-direction while the system effectively
extends to infinity along $x$ following the usual PBC (like in a nanoribbon).}

Our paper deals with systems of lattice size $L=6,~ 12,~18$ and 24 respectively. 
Due to numerical constraints, we use exact diagonalization only for smaller 6 and 12
site lattices while for $L=18$ and $24$, we use a modified Lanczos technique\cite{dag}.
This latter method search for the ground state {(a lowest quasi-energy state, in this case)} starting from a random state $\psi_0$,
with nonzero overlap with the ground state.
This is then acted upon by the Floquet Hamiltonian $H_F$ to obtain the state $\psi'_0=\frac{H_F\psi_0-<H_F>\psi_0}{\sqrt{<H_F^2>-<H_F>^2}}$, which is orthogonal to $\psi_0$.
The Hamiltonian, in its $2\times 2$ representation spanned by the basis states $\psi_0$ and $\psi'_0$, is then diagonalized.
The lowest eigenstate is renamed as $\psi_0$ and iterations are continued until the true minimum energy state is obtained.
See Ref.\onlinecite{dag} for details.

Given the state, we can now compute the concurrence between
NN or NNN spins.
Let's call the spin-$z$ basis vectors as $|\phi_j>$'s, in terms of which we can write the eigenstates of our Hamiltonian
as $|\psi_i>=c_{ij}|\phi_j>$ and let $E_i$ denote the $i$-th eigenvalue.
The ground state density matrix will then be given by $\rho_G=|\psi_0><\psi_0|$.
We can also compute the thermal density matrix which, in the canonical ensemble,
 is given by $\rho_T=\frac{1}{Z}\sum_ie^{-\beta E_i}|\psi_i><\psi_i|$, with $Z=\sum_ie^{-\beta E_i}$.

As the system is bipartitioned into subsystems $a$ and $b$, we can write $|\phi_i>=|\phi_i^a>\otimes |\phi_i^b>$ and the 
reduced density matrix in the subsystem $a$ will be given as
\begin{eqnarray}
 \rho^R_{i_aj_a}=\frac{1}{Z}\sum_{k,i_b,j_b}c_{ki}^\star c_{kj}e^{-\beta E_k}\delta_{\phi_i^b,\phi_j^b}.
\end{eqnarray}
Our subsystem $a$ consists of a pair of spins, especially NN or NNN pairs, that we consider here.
{We should mention here that our work involves bi-partite entanglement alone and does not deal with entangled states (like a GHZ state or a W state) corresponding to further partitioning of systems.}

Now let us look at the definition of quantum concurrence. For a 2-qubit system, the pure state $|\psi>$ contains a measure of 
concurrence $C(|\psi>)=|<\psi|\tilde \psi>|$, where $|\tilde \psi>$ is the time reversed state of $|\psi>$. For a spin system,
a time reversed state is the spin-flipped state and for a spin-1/2 (2-qubit) system it is given by $|\tilde \psi>=(\sigma_1^y
\otimes\sigma_2^y)|\psi^\star>$.
When we have a general mixed state, full information of the wave function is not available and the time-reversed density matrix is
obtained instead, as ${\tilde\rho}^R_{12}=(\sigma_1^y\otimes\sigma_2^y)\rho^{R^\star}_{12}(\sigma_1^y\otimes\sigma_2^y)$,
to compute the concurrence.
Here $\rho^R_{12}$ denotes the reduced density matrix at the reduced
2-qubit level. Concurrence becomes a function of $\rho^R_{12}$
and can be shown\cite{wooters} to be given by
\begin{equation}
 C(\rho)=max\{0,\lambda_1-\lambda_2-\lambda_3-\lambda_4\}
\end{equation}
where $\lambda_i$'s denote the square root of eigenvalues of $R_{12}=\rho^R_{12}{\tilde\rho}^R_{12}$
in descending order.

\begin{figure}[t]
\centering
\includegraphics[width=.98\linewidth,height=2.8 in]{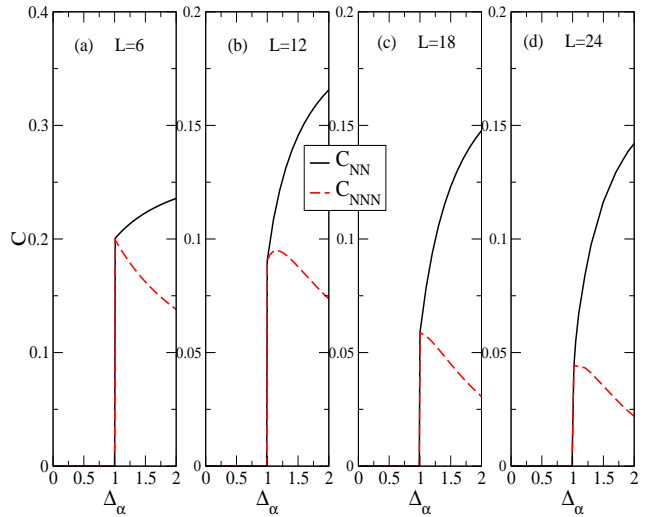}\
\caption{$C_{NN}$ and $C_{NNN}$ versus {$\Delta_\alpha$ at $\omega\rightarrow\infty$ limit} for $L=6,~ 12,~18$ and $24$ respectively.}
\label{fig2}
\end{figure}

\begin{figure}
\centering
\includegraphics[width=.9\linewidth,height=3 in]{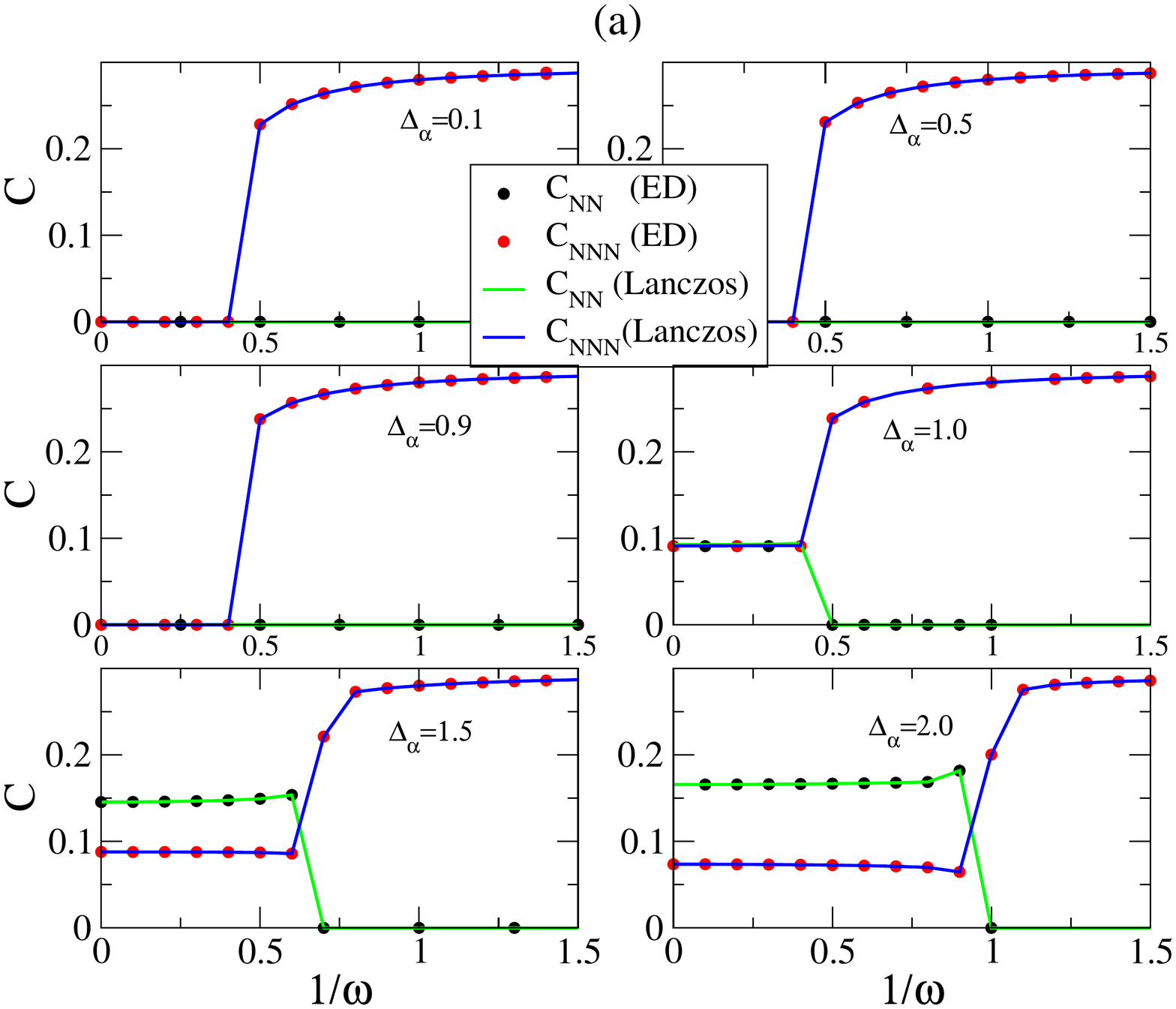}\\\vskip .05 in
\includegraphics[width=.9\linewidth,height=3 in]{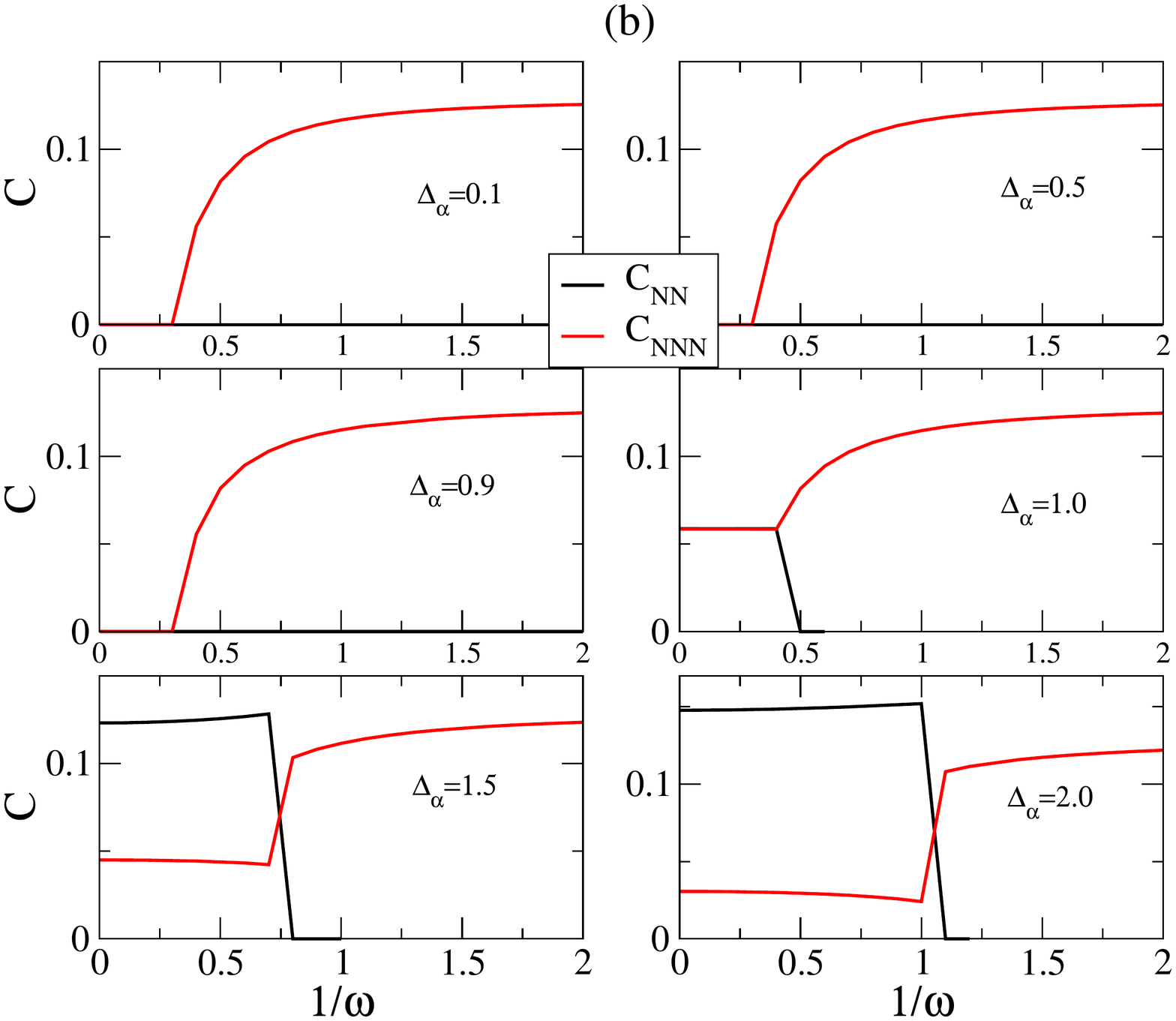}
\caption{Concurrence versus {$1/\omega$ (in units of $K(\alpha)$)} for(a) $L=12$ and (b) $L=18$.
In (b), results are obtained using modified Lanczos technique whereas in (a) both exact diagonalization 
(circles) and Lanczos (lines) results are shown.}
\label{fig3}
\end{figure}

\section{Results and Discussion}
First we study the effect of anisotropy or $\Delta_\alpha$ at $\omega\rightarrow\infty$ limit, $i.e.,$ the anisotropic
Heisenberg ferromagnetic spin system alone.
{Within the lowest quasi-energy state,} we probe the  concurrence between nearest neighbor sites ($C_{NN}$) and that between
next-nearest neighbor sites ($C_{NNN}$) {of the Floquet model}.
See Fig.\ref{fig2} for results of C vs. $\Delta_\alpha$ for lattices with $L=6,~12,~18$ and $24$.
{Physically, variation of $\Delta_\alpha$ can be achieved by varying $\Delta_0$, keeping $\alpha$ fixed.}
We see that concurrence becomes nonzero abruptly at $\Delta_\alpha=1$ signalling an entanglement transition.
{This feature remains intact in the thermodynamic limit as well, as shown
via finite size scaling in Appendix B.
However, the jump/discontinuity reduces as lattice size is increased.}

Next we consider the situation as the DMI is turned on, with decrease of $\omega$ to finite large values. We see the 
concurrence $C_{NNN}$ to appear and then increase gradually beyond some cut-off $D_\alpha(\omega)$ values for $\Delta_\alpha<1$ whereas $C_{NN}$ always remains zero. 
That cut-off remains the same as long as the Ising anisotropy remains. 
For $\Delta_\alpha\ge 1$, the system is already entangled at $D_\alpha(\omega)=0$. {There is also a cut-off  $D_\alpha(\omega)$ strength in this case, beyond which $C_{NN}$ perishes and $C_{NNN}$ shoots up and keep on increasing to reach a plateau finally.
We should mention here that for a fixed $\alpha$, $D_\alpha(\omega)$ is inversely proportional to $\omega$ and as such $D_\alpha(\omega)$ can be replaced with $\omega^{-1}$ to visualize the frequency dependence of the concurrences more clearly.  Accordingly,
Fig.\ref{fig3} shows variation of concurrences in terms of $\omega^{-1}$ in units of the prefactor $K(\alpha)$.}

Now let us explain the results of concurrence that we get.
At very large frequency, the DMI term becomes negligible and the XXZ model
of the Floquet Hamiltonian shows finite entanglement as soon as the anisotropy
becomes easy planar. Product state of the easy axis FM turns into a
entangled state with moments oriented in the spin-$xy$ plane\cite{karkeola}.
Finally for $\Delta_\alpha\rightarrow\infty$, the ground state still has no product state form 
as no direction in $xy$ plane is preferred for the spin-moments.
As a result, the entanglement, emerging from $\Delta_\alpha\rightarrow1+$, gradually saturates to a finite value
for large $\Delta$.
With {$\omega\rightarrow\infty$ (or, $D_\alpha(\omega)=0$),} there is no direct interaction between NNN spins.
Heisenberg point being the critical point, spin correlation is at its peak for closest spins, which then decays as 
the distance between the spins is increased. 
Hence both NN and NNN spins are very much correlated as well as entangled at $\Delta_\alpha=1$.
Moreover, as the entanglement producing spin-fluctuation terms appear only between NN spins, we find $C_{NN}> C_{NNN}$ 
whenever they are nonzero. 
We should emphasize here that as spin exchange between NN pairs entangles them more, we see $C_{NN}$ to increase
steadily with $\Delta_\alpha$ beyond the Heisenberg point.
This pushes $C_{NNN}$ for steady decrease possibly due to spin conservation {or the monogamy of entanglement\cite{bal}}.
Let us add here that the bump in $C_{NNN}$ observed for $\Delta_\alpha\rightarrow1+$ at $L=12$ is a finite size effect which
gets wiped off significantly in the plot corresponding to $L=18$ and $24$.
Notice that it does not appear for $L=6$ as PBC makes this a special case where the NN and 3rd NN sites often become identical.

\begin{figure}[t]
\centering
\includegraphics[width=\linewidth,height=2.5 in]{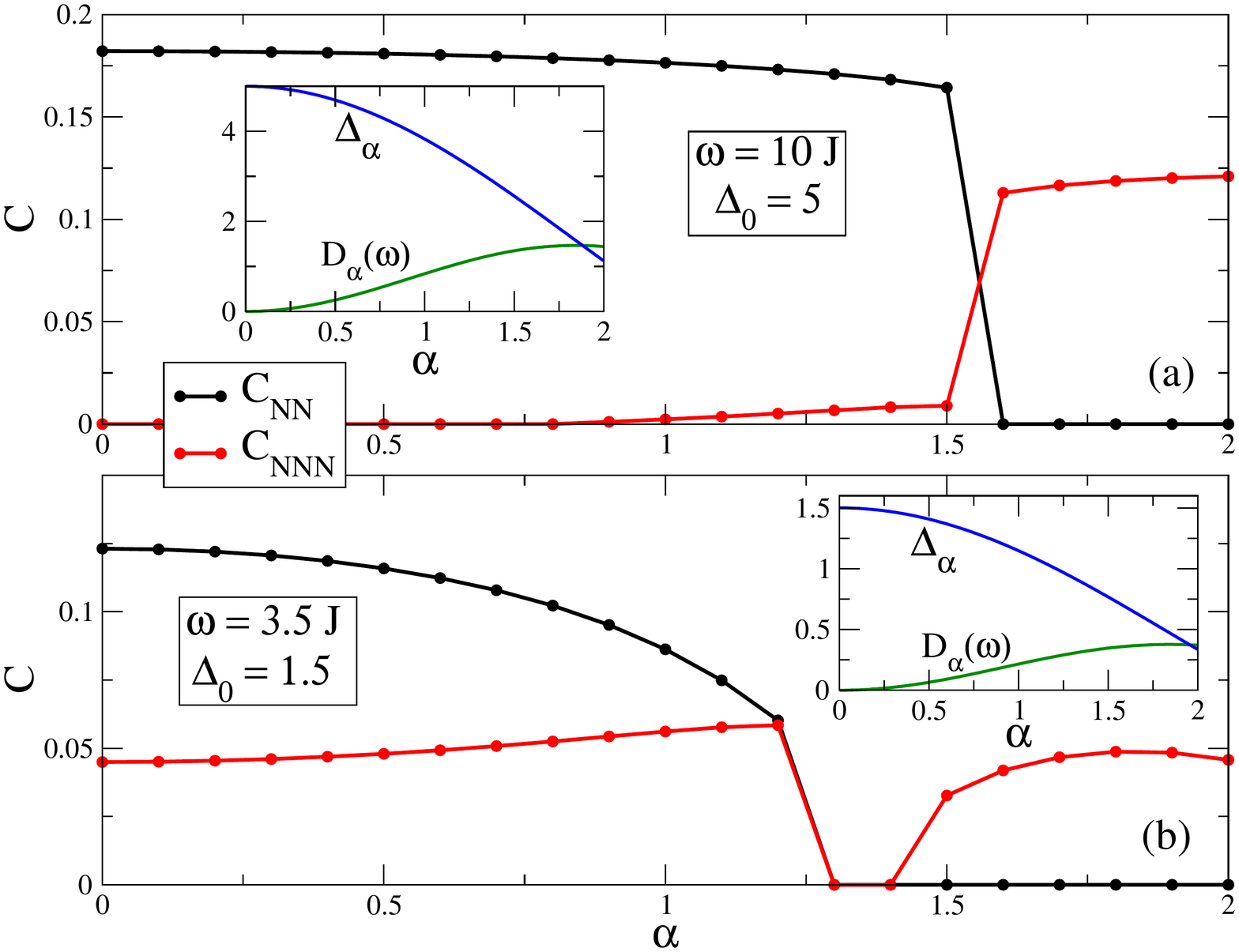}
\caption{(Color online) Variation of $C_{NN}$ (black) and $C_{NNN}$ (red/grey) as a function of $\alpha$ on a $L=18$ size lattice for (a) $\omega=10J~\&~\Delta_0=5$ and (b) $\omega=3.5J~\&~\Delta_0=1.5$. The inset shows the variation of $\Delta_\alpha$ and $D_\alpha(\omega)$ with $\alpha$.}
\label{alpha}
\end{figure}

Now as the DMI term is turned on, due to decrease of the frequency of irradiation from very large values,
the system becomes topological.
However, it takes some finite $D_\alpha(\omega)$ values to get the {Floquet} system with easy axis 
anisotropy to become entangled for NNN spin pairs.
This is because the NNN spin fluctuation terms oppose the FM ordering and it takes a finite threshold to disrupt that ordering and 
set in entanglement.
On the other hand the NN pairs never get entangled by introduction of this complex NNN hopping term.
The easy plane ferromagnet, which had finite $C_{NN}$ and $C_{NNN}$, shows decrease and increase in entanglement
with $D_\alpha(\omega)$ for NN and NNN pairs respectively.
The DMI term is a precursor of the spin-orbit coupling in the system and it
favors spin canting. As this term acts between NNN pairs,  a strong $D_\alpha(\omega)$
 indicates a larger correlation among the NNN pairs. But it also competes with NN spin exchange term and let
 the correlation between the NN pairs perish gradually.
Fig.\ref{fig3}(a) shows the concurrence results for $L=12$ where both exact diagonalization and Lanczos results are shown
which fairly matches for the values of anisotropies considered. For larger $L=18$ site lattice, we use Lanczos
method and obtain same qualitative results as shown in Fig.\ref{fig3}(b).
Notice that for easy planar case, $C_{NNN}$ shows two smooth branches connected by a jump/discontinuity in the middle {(however, such high frequency branch vanishes for very large $\Delta_\alpha$, as also can be seen in Fig.\ref{alpha}(a) which shows $C_{NNN}=0$ for small $\alpha$'s). Using finite size scaling analysis, we have seen this to exist even in the thermodynamic limit (see Appendix B).
This is a phase transition in which a redressing of the spins develop within the spin-$xy$ plane.
  An easy planar ferromagnet already has entangled NN and NNN spin-pairs even without any DMI term. With finite $\omega$, DMI is brought in which 
  opposes the existing NNN spin ordering (and that enhances $C_{NN}$ accordingly due to monogamy of entanglement) resulting in slight reductions in $C_{NNN}$ with $\omega$.
On the other hand, the low $\omega$ or large $D_\alpha(\omega)$ branch of the plot
appears beyond the cut-off $D_\alpha(\omega)$ strength, like in the easy axis case, and characterizes increase in NNN spin correlation (in its DMI induced new spin ordering) with $D_\alpha(\omega)$ or $\omega^{-1}$.}
\begin{figure}[htb]
\centering
\includegraphics[width=\linewidth,height=3 in]{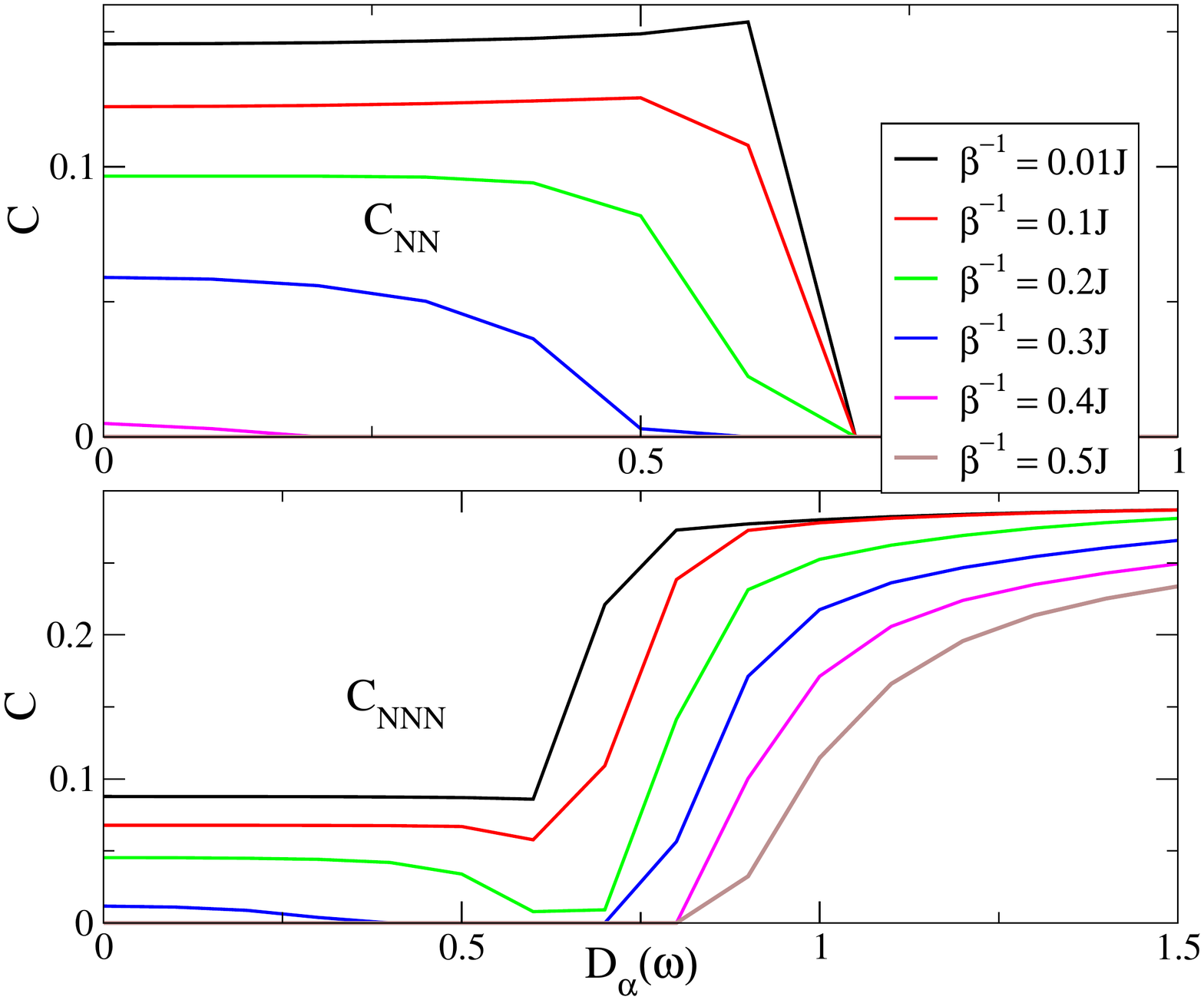}\\
\vskip .1 in
\includegraphics[width=.49\linewidth,height=2 in]{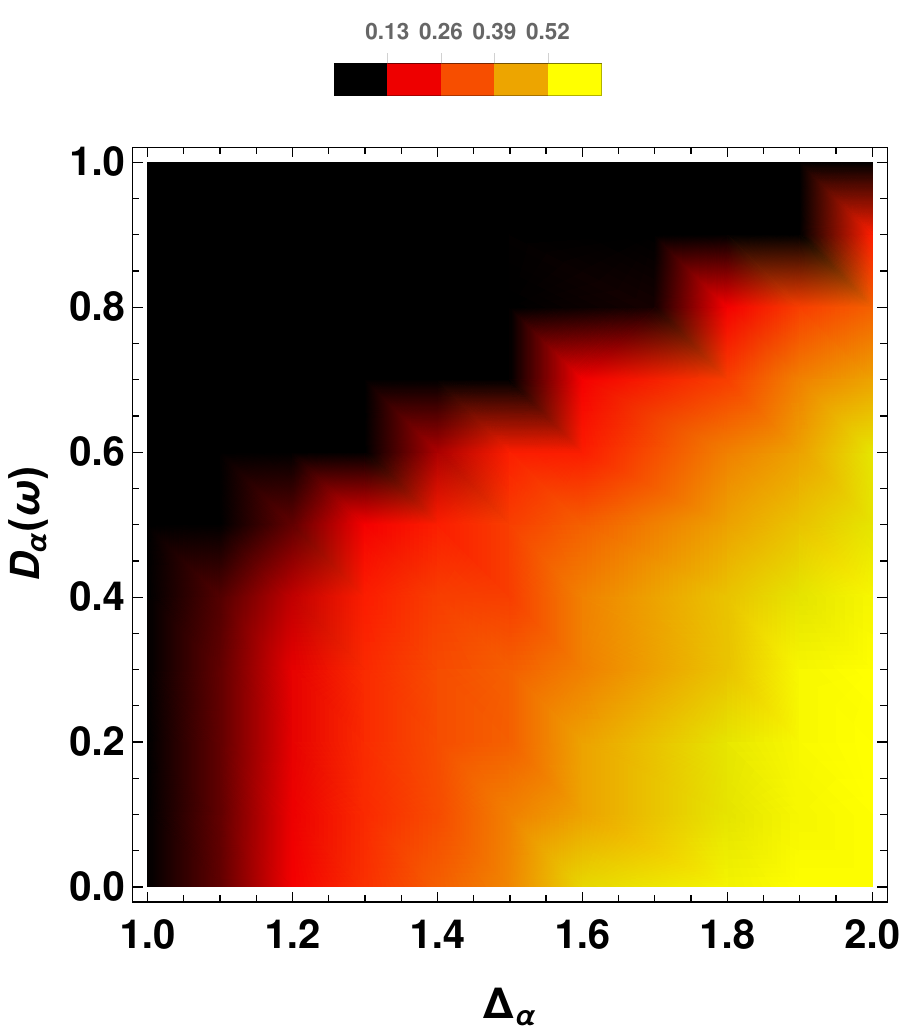}~~
\includegraphics[width=.49\linewidth,height=2 in]{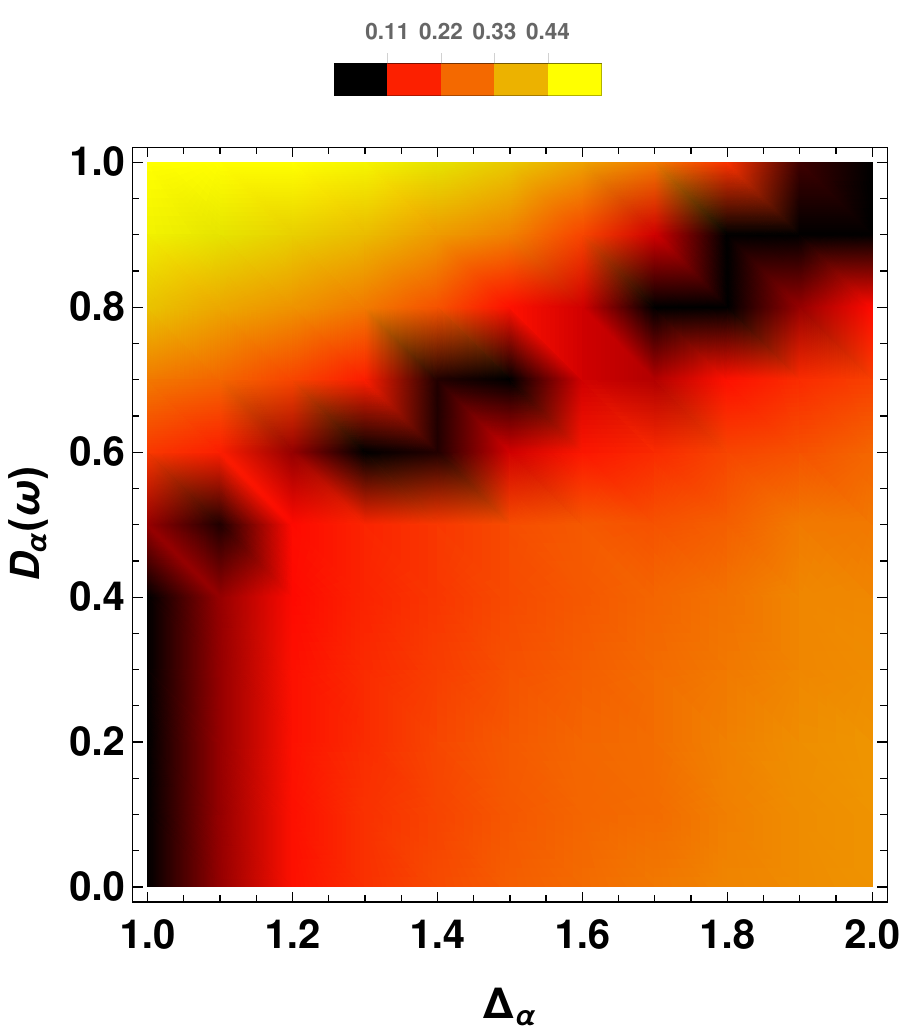}
\caption{(top)Thermal Concurrence for $L=12$ at $\Delta_\alpha=1.5$. Threshold temperature $T_{th}$
for $C_{NN}$ (bottom-left) and $C_{NNN}$ (bottom-right) in the $D_\alpha(\omega)-\Delta_\alpha$ plane.}
\label{fig4}
\end{figure}

{Our results on $C$-vs-$1/\omega$ plots describe variation of concurrences as the $D_\alpha(\omega)$ is varied, keeping $\alpha$ fixed. But it is also useful to look at the variation of concurrences with $\alpha$ (see that $\alpha$ is proportional to the electric field amplitude $E_0$) for fixed large $\omega$ values.
In Fig.\ref{alpha} we can see such variations for two different sets of $(\omega,\Delta_0)$. In Fig.\ref{alpha}(a) and within the range of $\alpha$ shown, easy planar anisotropy is experienced by the Floquet system. It shows that a large anisotropy $\Delta_\alpha$ can push $C_{NNN}$ to zero even at the unirradiated limit $\alpha=0$. In Fig.\ref{alpha}(b), we consider a comparatively small $\omega$, yet being large compared to $J$ and $\Delta_0J$.
It shows transition from easy planar to easy axis anisotropy (see the inset).
With increase of $\alpha$, the anisotropy $\Delta_\alpha$ changes from easy planar to easy axis type (beyond $\alpha=1.2$)
and that makes $C_{NN}$ to go to zero for all larger $\alpha$ values, whereas the behavior of $C_{NNN}$ demonstrates aptly the sudden death and revival of entanglement\cite{sudden-death} as it remains zero only if the $D_\alpha(\omega)$ is less than the cut-off value as mentioned in discussion pertaining to Fig.\ref{fig3}.}

\subsection{Thermal Entanglement}

We know that the ground state is realized at zero temperature and in practice,
low energy excitations also need to be taken into account to understand the low temperature phenomena in a system.
Thus in our case, its wise to take a look at the thermal entanglement that corresponds to entanglement properties at a finite temperature.
Fig.\ref{fig4} shows the results of thermal concurrences $C_{NN}$ and $C_{NNN}$ in a $L=12$ size system obtained
for various $\beta~(=\frac{1}{k_BT})$ values, and for $\Delta_\alpha=1.5$.
Our thermal entanglement results show that temperature causes the entanglement measure in the system to wear off and with high 
temperature, the thermal fluctuation leads the system towards complete unentanglement. As further quantification, we compute
the threshold temperature $T_{th}$ above which there is no concurrence possible in the Floquet states.
In bottom panel of Fig.\ref{fig4}, we show the variation of $T_{th}$ for $C_{NN}$ and $C_{NNN}$ in a $D_\alpha(\omega)-\Delta_\alpha$ plane. 
It shows that a large $\Delta_\alpha$ ($i.e.,$ much larger than unity) keeps the NN spins entangled upto some appreciably large $T_{th}$
values, if the irradiation born $D_\alpha(\omega)$ term is not very strong. On the other hand, a large $D_\alpha(\omega)$
makes the NNN spins entangled with appreciably large $T_{th}$ values when easy planar $\Delta_\alpha$ is not very large.
\begin{figure}
\centering
\includegraphics[width=\linewidth,height=2. in]{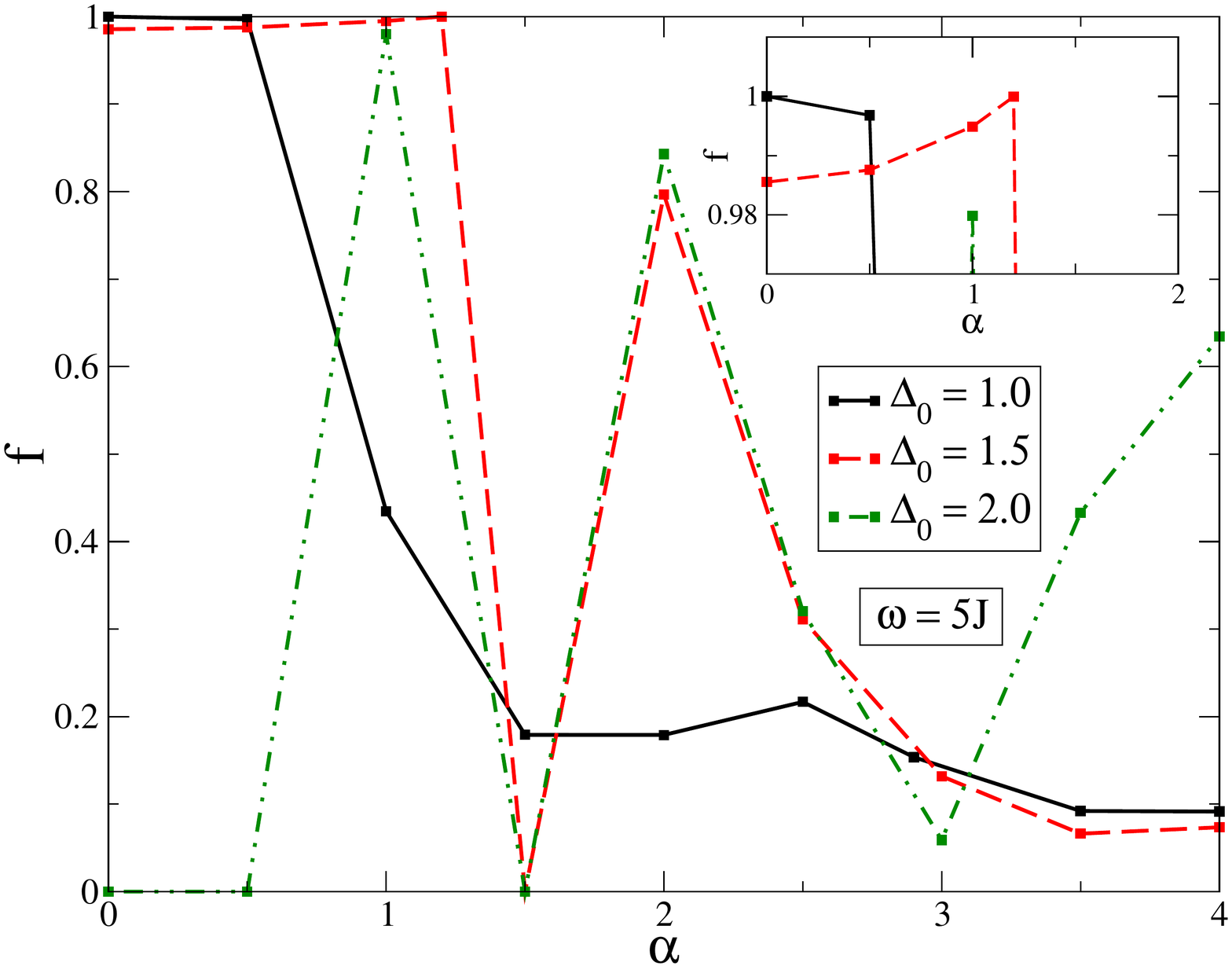}
\caption{Variation of fidelity of the lowest quasi energy state between the edge-less and edged configurations corresponding to $\Delta_0=1.0,~1.5~\&~2.0$ and $\omega=5J$ at $L=18$. The inset shows the same plot zoomed in around f=1.}
\label{fid}
\end{figure}
\begin{figure}[b]
\centering
\includegraphics[width=\linewidth,height=3 in]{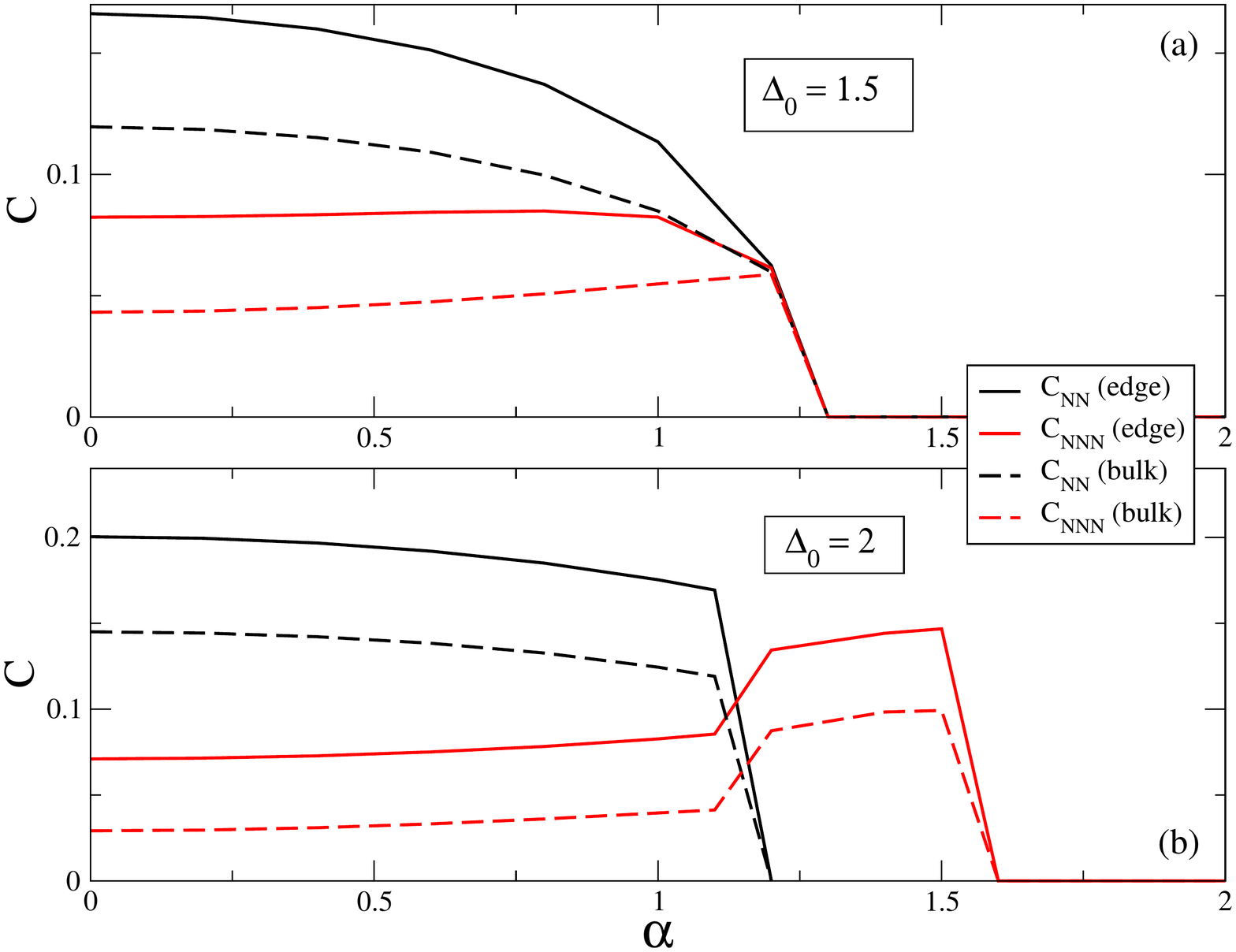}
\caption{($C_{NN}$ (black) and $C_{NNN}$ (red/gray) versus $\alpha$  at the edges (solid lines) and within the bulk (dashed lines) of a $L=18$ size lattice with PBC along $x$ and OBC along $y$ for $\omega=5J$ and for (a) $\Delta_0=1.5$ and (b) $\Delta_0=2.0$ respectively.}
\label{edge}
\end{figure}
\begin{figure}[t]
\centering
\includegraphics[width=\linewidth,height=2.5 in]{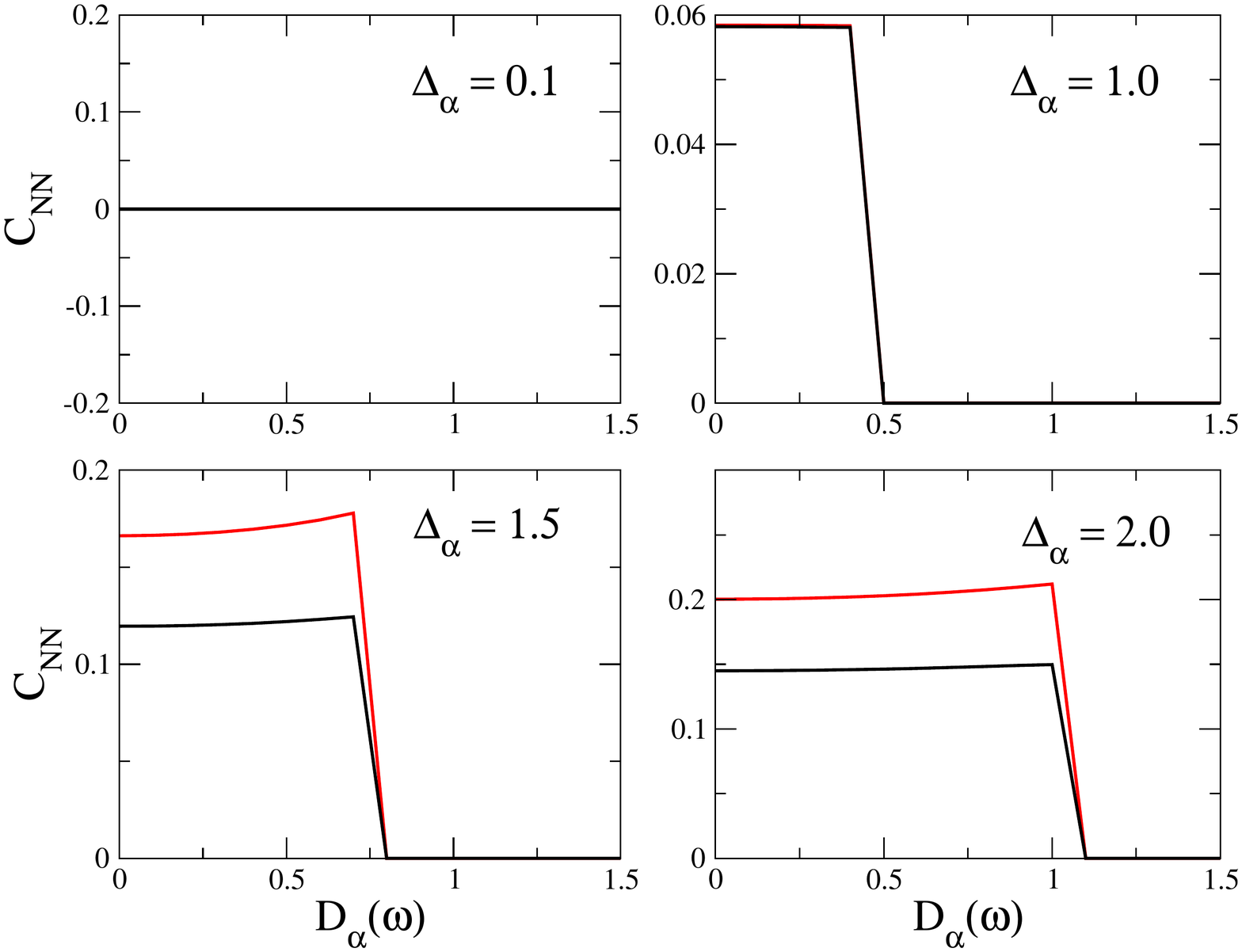}\\
\includegraphics[width=\linewidth,height=2.5 in]{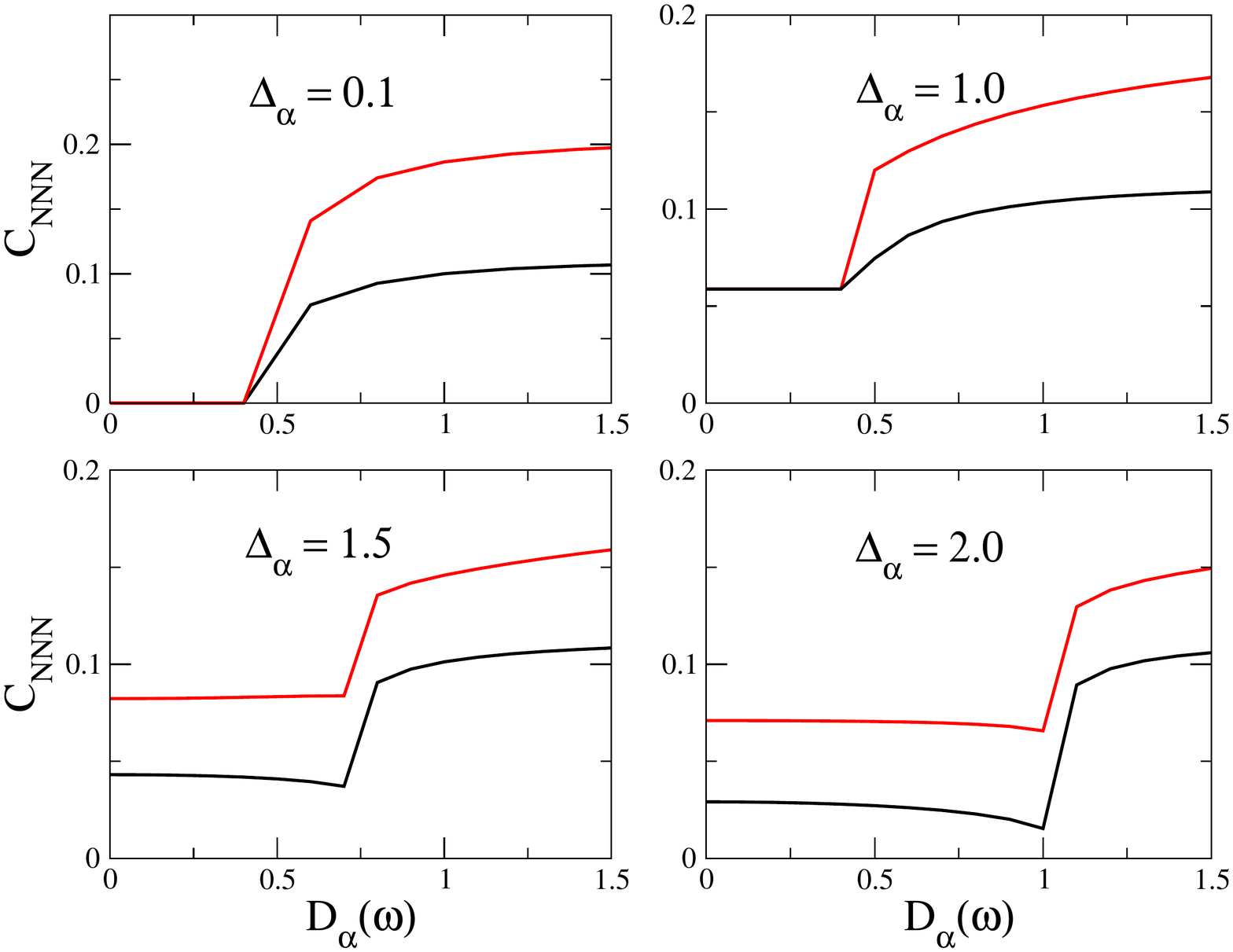}
\caption{(top)$C_{NN}$ and (Bottom) $C_{NNN}$ versus $D_\alpha(\omega)$, for fixed $\Delta_\alpha$ values, within the bulk (black lines) and at the edges (red lines) on a $L=18$ size lattice with PBC along $x$ and OBC along $y$.}
\label{edge2}
\end{figure}

\subsection{Results for finite geometries with edges}
{As the presence of the DMI term brings in topological nontriviality to the Floquet problem, we need to pay special attention to the edges.
  Hence we consider finite size clusters with PBC along $x$ and OBC along $y$ so as to produce nano-ribbon geometries with zig-zag edges at the top and bottom along $x$ direction and try to quantify the edge correlations in the system. 
  First we calculate the fidelity $f$ ($i.e.,$ wave-function overlap) between the lowest quasi-energy states of an edge-less (with PBC along $x$ and $y$) and edged (with PBC along $x$, OBC along $y$) system (for L=18) as $H_F$ is optically tuned keeping $\omega$ fixed.
  A few results are shown in Fig.\ref{fid} for isotropic as well as easy planar configurations. The oscillating behavior appears due to the presence of Bessel's functions within the Hamiltonian parameters.
  At Heisenberg point, we witness $f=1$ in the unirradiated limit, implying identical states for the edged and edge-less configurations. Hence, like any property, entanglement measures also do not change by merely bringing in such edges. But larger anisotropies (like $\Delta_0=1.5,~2.0$, as shown in Fig.\ref{fid}) causes states to differ even in the unirradiated limit and we get different measures for concurrences for edge spin pairs and bulk spin pairs (see Fig.\ref{edge}). We should mention here that this fidelity calculation can only indicate identical or non-identical lowest quasi-energy state entanglement behaviors depending on whether $f=1$ or $f\ne1$. For $\alpha\ne0$, we obtain $f\ne1$ in general and different entanglement measures can be expected for the edged configuration. As long as $\Delta_\alpha=1$, our results 
   show full fidelity (see Fig.\ref{fid} and the corresponding entanglement match in Fig.\ref{edge}). We can say that for those cases, edge states 
  are not present in the lowest quasi-energy states of the FTI. Other than those points, we witness both $C_{NN}$ and $C_{NNN}$ to reach larger values at edges as compared to that in bulk, in the nanoribbon geometry considered
  (see Fig.\ref{edge}-\ref{edge2}). 
  Here we look for one possible explanation of such behavior. These points corresponds to $f<1$ and thus the wave-functions at the edges are more likely to differ from that from an edge-less configuration. So we can say that the lowest quasi-energy state do contribute to the edge states, which indicates gaplessness of the spectrum. 
  Now we know that the entanglement entropy of a short-ranged gapped system show areal law behavior for the ground state entanglement\cite{hastings} whereas for a gapless system a logarithmic correction is added
to that with prefactor proportional to the central charge of the corresponding conformal field theory at the critical point\cite{cala}. This makes entanglement at the gapless point to be higher than that of a gapped regime.
We find that the entanglement measure of two qubit concurrence, that we calculate here, also demonstrates similar behavior and produces larger concurrences at the edges than within the bulk.}

\section{Conclusion}
In this work, we have studied spin-spin entanglement in a Floquet system arising out of a FMHS model in a honeycomb lattice
irradiated via circularly polarized light.
{Though this work can be termed as a simple study of entanglement for a spin-1/2 XXZ model with NNN DMI on a honeycomb lattice, the easy synthetic tunability of the Floquet system makes this work stand out firmly of the rest for we have the freedom to adjust the parameters of the Hamiltonian.
We find that just by varying the amplitude and frequency of irradiation, and not directly modifying the anisotropy or DMI strengths as such,} can lead to a plethora of interesting 
findings,
in 2-spin ground-state as well as thermal concurrences.
{Firstly, when $\omega$ is very large (as compared to $J$ and $\Delta_0J$), the DMI contribution is negligible and 
increasing the field strength reduces the spin anisotropy $\Delta_\alpha$.
For the easy planar scenario, reduction of $\Delta_\alpha$ comes with decrease (increase) of $C_{NN}~(C_{NNN})$. This occurs as  $\Delta_\alpha$ quantify interactions between NN spin-pairs as well as due to the monogamy of entanglement.
Across the Heisenberg point corresponding to the Floquet model, a transition develops from entangled to unentangled NN and NNN spin pairs.}
If the original spin anisotropy $\Delta_0$ is barely above unity, high frequency irradiation can make the system
unentangled producing separable product states in the lowest energy eigenfunctions.
 
Now as the frequency becomes intermediate so as to make $D_\alpha(\omega)$ appreciable, the system becomes topological. We see no 
coincidence between topological and entanglement transitions occurring there.
In fact, this is not surprising as our working model obtained
from the Floquet theory, comprises of short-range interaction/spin fluctuations
alone and hence unlike in long-range entangled fractional Hall systems\cite{maria}, we don't see any immediate
entangling or disentangling as the DMI term is turned on. 
However, we notice interesting nontrivial entanglement features in presence of the DM term.
The easy axis Floquet FMHS system produces non-zero $C_{NNN}$ beyond a cut-off $D$ value, as the DM term competes with
the NN spin flip term of the Hamiltonian.
For the easy-planar case, both $C_{NN}$ and $C_{NNN}$ are nonzero without a DM term. Here also $C_{NNN}$ shoots up to a
higher value beyond a cut-off $D_\alpha(\omega)$ while $C_{NN}$ reduces down to zero.
{When we vary the field strength (which is proportional to $\alpha$),
we find that we can choose to have convenient parameters so that sudden-death and revival/rebirth of entanglement can be observed. This is due to transition of $\Delta_\alpha$ between easy axis and easy planar type which show different entanglement behaviors.}
Furthermore, we study thermal concurrence to demonstrate how system entanglement steadily decreases with the temperature.

{Other than the bulk, we also study a zig-zag edged configuration and probe the effect of high frequency irradiation on that.
We find that the lowest quasi-energy state differs due to the development of edges in the easy planar Floquet system and causes the concurrence measures to be higher at the edges as compared to that within the bulk. }

Ours is an important piece of work as the concurrence patterns obtained can be useful in extracting quantum information from 
various QMBS. For example,
controlled creation or destruction of entanglement via tuning concurrence of the Floquet states has already been shown for periodically driven coupled flux qubits\cite{sanchez}.
Structures of entanglement for both surface and bulk states are 
 examined in the
topological insulator $Bi_2Te_3$\cite{panigrahi} or the
full density matrix of two qubit systems have been measured experimentally and the corresponding concurrence and fidelity computed\cite{shulman}.

Our thermal concurrence results also add important insight to the low temperature
entanglement behavior in QMBS.
Down the line, one can also  explore
the effect of transverse (normal to easy direction) magnetic field on the spins that 
sometimes witness enhancement of thermal entanglement with temperature\cite{jafari} (which is not the usual behavior).
Besides, it will also be interesting to 
quantify the quantum coherence\cite{chand} or perform the 
{Bell-state measurement\cite{bell} on the entangled Floquet states.} In short, we believe that the present study may trigger various
further analytic as well as experimental researches with possible connection to spintronics and topological computations.
{We should mention here that experimentally, quantification of concurrence is possible using different protocols
for both pure states\cite{entropy,measureC} and mixed states\cite{entropy} and it will be interesting if that can be pursued for our Floquet system and compared with our numerical results.}

\begin{acknowledgements}
SK thanks S. Ghosh, S. Mandal, S. Yarlagadda and K. Sengupta for valuable comments, acknowledges CSIR, India for the financial support
and IACS, India for providing computational facilities.
Both the authors thank  S. Owerre and D. Chowdhury for useful discussion on the work. Besides, the authors also acknowledge the anonymous referees for their comments and suggestions that resulted in considerable improvement of the paper.
\end{acknowledgements}

\appendix
\section{Floquet Hamiltonian}
In a honeycomb lattice, there are
two sublattices designated by A and B. Each of those sublattices can be defined using three unit vectors ${\bf e_1}=(0,a),~{\bf e_2}=
(-\frac{\sqrt{3}}{2}a,-\frac{a}{2}),~{\bf e_3}=(\frac{\sqrt{3}}{2}a,-\frac{a}{2})$. Let us consider $a$, the length of
the NN bonds to be unity. Fig.1a shows a cartoon of the same.

A ferromagnetic XXZ spin-$1/2$ model is given by the Hamiltonian 
\\ $H=-\sum_{<\alpha,\beta>}[JS_\alpha^zS_\beta^z+\frac{J_{\perp}}{2}(S_\alpha^+S_\beta^-+h.c.)]$.\\\\
Under a Holstein-Primakoff transformation, this takes the form:
$H=\sum_k\psi_k^\dagger H_k\psi_k$ with
$\psi_k=(a_k,b_k)^T$ and $H_k=3JS[\sigma_0-\Delta(\sigma_+\gamma_k+h.c.)]$. Here $\Delta=J_\perp/J,~\sigma_+=(\sigma_x+i\sigma_y)/2$ and 
$\gamma_k=\frac{1}{3}\sum_j e^{-ik.e_j}$. $a_k,~b_k$ denote the magnon annihilation operators and $\sigma_i$'s are Pauli matrices
to describe the pseudospins.
The energy dispersion becomes $\epsilon_k=3Js(1\pm\Delta|\gamma_k|)$ that gives degeneracy at
the Dirac points $K_\pm=(\pm\frac{4\pi}{3\sqrt{3}},0)$. Also notice that the Dirac nodes appear with nonzero energy $3JS$.

Upon irradiation via circularly polarized light with $E=E_0 (\tau cos(\omega t),sin(\omega t))$ (with $\tau=\pm 1$), 
an additional phase is added, due to Aharonov-Casher effect, to the amplitude of the spin fluctuation term involving
site $i$ and $j$:
\begin{equation}
 \phi_{ij}=\frac{1}{\hbar c^2}\int_{r_i}^{r_j}\boldmath E\times\mu.dl
\end{equation}
where spin moment {\boldmath$\mu$}$=g\mu_B{\hat z}~(g$ and $\mu_B$ are gyromagnetic ratio and Bohr magneton respectively). This brings in the time dependence as 
\begin{eqnarray}
 H(t)=-J\sum_{<i,j>}[S_i^zS_j^z+\frac{\Delta }{2}(e^{i\phi_{ij}}S_i^+S_j^-+h.c.)]
\end{eqnarray}

For studying dynamics using Floquet theory, first
the Fourier components of the Hamiltonian are obtained and they are given as 
\begin{eqnarray}
H^{(n)}&=&\frac{1}{T}\int_0^Tdt e^{-in\omega t}H(t)\nonumber\\
&=&-J\sum_{<i,j>}[\delta_{n,0}S_i^zS_j^z+\frac{C_n}{2}e^{-in\theta_{ij}}(S_i^+S_j^-+h.c.)]\nonumber\\
\end{eqnarray}
 where $C_n={\rm J}_n(\alpha)\Delta$ and $\theta_{ij}$ denotes the angular orientation of the $(i,j)$ bond.\\
 For large $\omega$, we utilize a high frequency expansion which gives an effective stationary Hamiltonian
 to the problem: $H_{eff}=\sum_i H_{eff}^{(i)}/\omega^i$.
 
For the present case, we obtain $H_{eff}^{(0)}=H^{(0)}=$\\
$-J\sum_{<i,j>}[S_i^zS_j^z+\frac{C_0}{2}(S_i^+S_j^-+h.c.)]$
 and \\$H_{eff}^{(1)}=\sum_{n=1}^\infty  \frac{1}{n}[H^{(n)},H^{(-n)}]$.
 This 1st order correction turns out to be 
 $H_{eff}^{(1)}/\omega=D_F\sum_{ij~\rm pairs}\nu_{ij}S_k.(S_i\times S_j)$.
Here $\nu_{ij}=+1~(-1)$ for $i,j\in~A(B)$ sublattice and $D_F=\sqrt{3} J^2C_1^2/\omega$.
{Thus, as long as $D_F$ is not negligible, compared to unity or $C_0/2$ ($i.e.,$ the strength of the two terms of $H^{(0)}$), we should consider this 1st order correction to the Floquet Hamiltonian.}
Similar calculations can be seen in Ref.~\onlinecite{eckart,owerre} as well.
{Lastly, we want to add here that the 2nd order correction is proportional to $\omega^{-2}$ and if we were to use Eq.\ref{eq2} for $H_F$, we must be careful not to choose $\omega$ small enough that this term also becomes non-negligible.}
\begin{figure}[t]
\centering
\includegraphics[width=\linewidth,height=3 in]{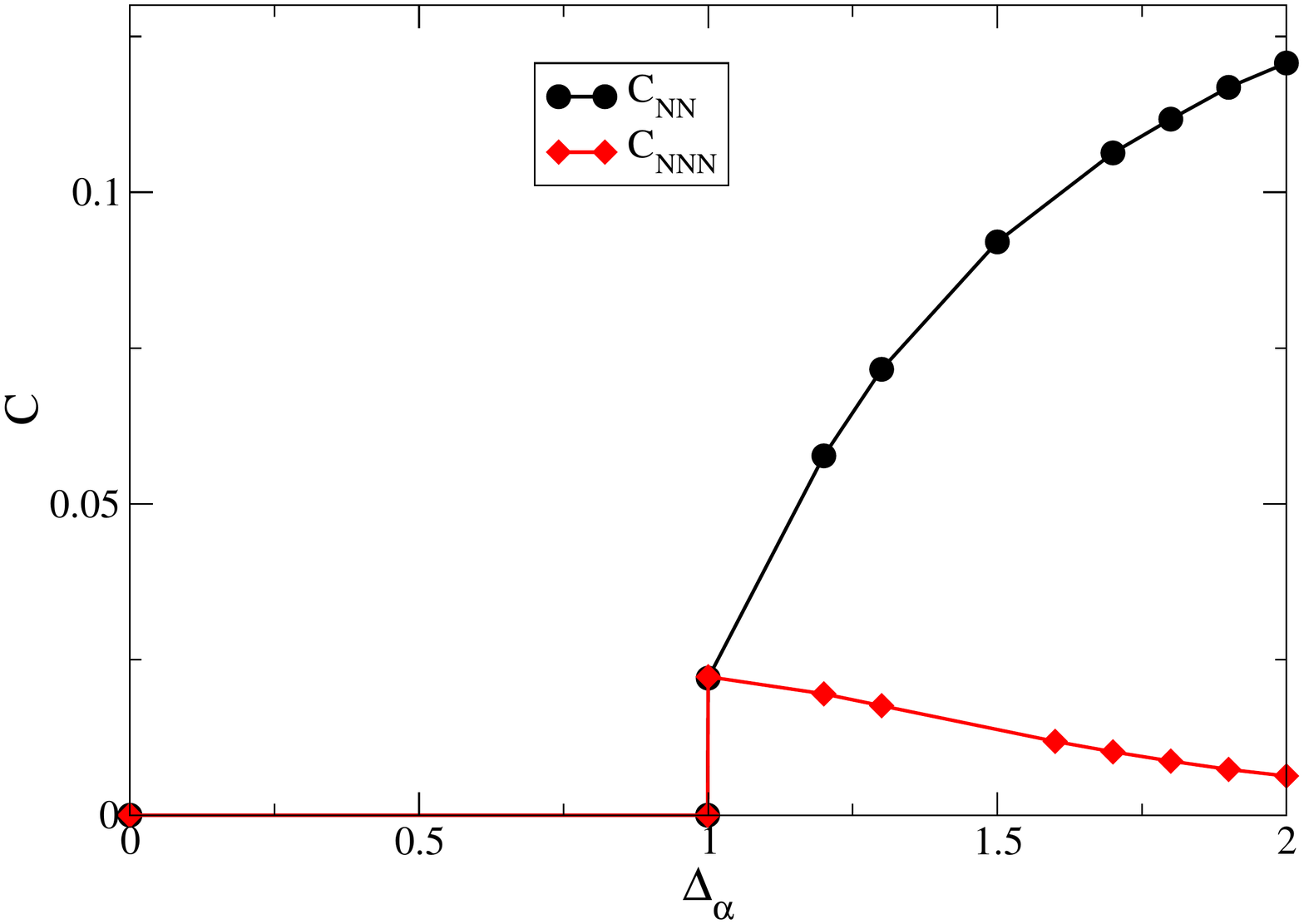}
\caption{Finite size scaling results: Asymptotic values of $C_{NN}~\&~C_{NNN}$ at $\omega\rightarrow\infty$ limit.}
\label{fss}
\end{figure}

\section{Finite size scaling}
{We did a finite size scaling analysis for concurrences at $\omega\rightarrow\infty$ limit involving $L=12,~18$ and $24$ size lattices which shows that the basic feature remains the same other than reducing the absolute values of $C_{NN}$ and $C_{NNN}$ to non-zero smaller values (see Fig.\ref{fss}). It is not possible to
do such analysis in presence of DMI, as $D_\alpha(\omega)$ itself shows some size dependence. But we can do finite size scaling analysis for the discontinuous jumps observed in Fig.\ref{fig3} and our calculations show its values to be 0.012 and 0.020 for $\Delta_\alpha=1.5$ and $2.0$ respectively.
This shows that such jump indeed exist in the asymptotic limit.}

\end{document}